\documentclass[12pt]{article}
\begin{document}

\title{\bf 6-D model with scalar field condensation at brane
and zero 4-D Cosmological Constant}

\author{E.I. Guendelman \\
\small\it Department of Physics,\\[-1.mm]
\small\it Ben-Gurion University, Beer-Sheva 84105, Israel  \\[-1.mm]
\small\it email: guendel@bgumail.bgu.ac.il}
\date{ }

\maketitle

\begin{abstract}
In a previous publication we have shown that the gauge theory of
relativistic $3$-Branes can be formulated in a conformally
invariant way if the embedding space is six-dimensional. The
implementation of conformal invariance requires the use of a
modified measure, independent of the metric in the action. We here
generalize the theory to include conformal invariance breaking and
a dynamical scalar field with a non-trivial potential. The non
conformal invariance contribution can be interpreted as
originating from a continious "non ideal brane fluid" that exists
between two singular branes. The scalar field potential also
breaks the conformal invariance. At singular brane locations,
conformal invariance is restored and the dynamics of the scalar
field is frozen at a certain fixed value of the scalar field which
depends on an arbitrary integration constant. Spontaneous Symmetry
breaking can take place due to such boundary condition without the
need of invoking tachyonic mass terms for the scalar field. In
these Brane-world scenarios, zero $4$-$D$ cosmological constant is
achieved without the need of invoking a fine tuned cosmological
constant in $6D$. Thus, no ``old'' cosmological constant problem
appears. The use of a measure independent of the metric is crucial
for obtaining all of the above results.
\end{abstract}

\newpage
 {\bf I Introduction}

In recent years a great deal of work has been done on the notion
that extended objects could play important roles in particle
physics and cosmology. In the context of string theory for
example, among  the various kind of branes a unique role is played
by $D$--branes [1] as they can trap the end-points of open
strings. $D$ -branes fit quite nicely with the idea, being studied
since the 80's, that our universe contains one or more branes
embedded in some higher dimensional space. These
``brane--universe'' models, are currently under investigation as
there is the hope that they will be of use in the solution of l
ongstanding hierarchy problems in gauge theories.\\
The gauge theory formulation of $p$--branes, proposed some years
ago as an alternative to the standard description of relativistic
extended objects [2],[3], is well suited to describe this new type
of cosmological scenario. Furthermore, the description of
$p$--branes in terms of associated gauge potentials offers a
vantage point to study some specific problem as the one
concerning the fine tuning of the cosmological constant.\\
We have shown [4] that for $3$-branes considered in an embedding
$6D$ space the gauge theory formulation of $3$-branes allows a
conformally invariant realization. An essential element necessary
to implement conformal invariance is the introduction of a measure
of integration in the action which is independent of the metric
[5],[6],[7],[8]. We use then such a formulation to construct a new
type of brane
world scenario.\\
Brane world scenarios in general are concerned with the
possibility that our universe is built out of one or more
$3$-branes living in some higher dimensional space, plus some bulk
component, [9],[10],[11],[12], [13]. In particular, the possibility of
$3$-branes embedded in $6D$ space has been studied in [14],
[15],[16],[17],[18]. In this case the effect of the tension of the
branes is to induce curvature only in the extra dimensions. In
these models there is still a question of fine tuning that has to
be addressed, since although the branes themselves do not curve
the observed four dimensions, the bulk components of matter do,
and they have to be fine tuned in order to get (almost)zero four
dimensional vacuum energy. This very special feature of $3$-branes
is a $6D$ embedding spacetime is related to the fact that such
matter content, even coupled to gravity, has a conformal
invariance
associated to it.\\
In our previous publication [4], in order to solve this problem,
we incorporated the ``brane-like features'' that are quite good in
what concerns the cosmological constant problem into the ``bulk''
part of the brane scenario as well. In this way both bulk and
singular brane contributions  shared the fundamental
feature of curving only the extra dimensions. \\
\newpage

Indeed, in Ref.[4],  we saw that the  GFF3B6D allows us to
understand , extend and give a ``pure brane interpretation'' of
the results of [19], where a ``square root gauge theory'', coupled
to $3$-branes in $6D$ was considered. This model has conformal
invariance and there is no need to introduce a $6D$ cosmological
constant. The ``fundamental physics'' behind the model was not so
clear however and its different matter elements: gauge fields,
$3$-branes appear rather disconnected from each other. Now with
the bulk being an 'ideal fluid of branes, the bulk and singular
brane appear in a unified framework. We showed in [19] that
GFF3B6D allowed us also  to interpret a more general set of
brane-world solutions, where the solutions
presented in [19] appear as very particular cases. \\
The objective of this paper is to show how the above picture
survives, even in the presence of conformal breaking terms. This
is of course very important sice our physical universe does not
satisfy the requirements of conformal invariance, since the
existence of massive particles is evidence against such a
symmetry. However, even departing from the conformally invariant
version of the GFF3B6D, we can mantain the basic feature that the
four dimensional part of the manifold does not get curved,
provided the use of the measure of integration independent
of the metric is mantained.\\
Other aspects or 'principles' can be relaxed. In the first place
we allow conformal invariance breaking which we introduce, to
demonstrate the basic mechanisms that the theory provides (other
generalizations are likely to produce similar results), in two
different ways: First in the GFF3B6D, we allow for 'non ideal' or
'interacting' behavior in the fluid of 3 branes (between the
singular branes) and second, we introduce a scalar field, with a
non trivial potential which will break also the conformal invariance.\\
As we will see, singular branes, as opposed to a continious
distribution of 3-branes, is necessarily conformal invariant.
Beyond the good behavior of the theory in relation to the
cosmological constant problem, another, rather amazing effect
appears here: it turns out that when considering a braneworld
where say two singular branes are considered and the space in
between is filled with 'non ideal' or interacting fluid of branes
plus a scalar field, one finds then that at the brane itself, the
scalar field is frozen at a particular fixed value, determined by
an arbitrary integration constant. This brane is indeed a $D$
brane, not just speaking from the possible underlying string
theory, but also from the point of view of the scalar field
expectation value, which gets fixed at a certain constant value in
the singular branes.\\
This effect makes it possible for the scalar field to break
spontaneously the symmetry, without having to relly on some
tachyonic mass.\\
The paper is organized as follows. In Sect.II, we review the
GFFB6D and and display its conformal invariant formulation when a
modified measure is introduced, also the dual picture to this
formulation is introduced; in Sect.III, the equations of motion in
this dual picture are studied. In section IV we study the
generalized GFFB6D, which includes conformal symmetry breaking
introduced through a 'non ideal' brane fluid behavior in a space
between singular branes and a scalar field with a non trivial
potential and how such formulation introduces Dirichlet boundary
conditions for the scalar field in the singular brane, while the
four dimensional space still does not get curved. We end up with a
brief discussion  and conclusions.\\
\newpage
{\bf II Conformally Invariant Realization in $6D$}\\
In a previous publication [4] we have shown that a fluid of
$3$-branes interacting with gravity can be formulated in a
conformally invariant fashion provided the embedding space is
$6$-D. The relevant action is (for a full treatment see ref. [4])

\begin{eqnarray}
&&S=-\frac{1}{16\pi G_{(5+1)}}\int d^{5+1}x \, \Phi\,
g^{{}^{AB}}\, R_{{}_{AB}}\left(\, \Gamma\,\right) +e^2\,\int
d^{5+1}x \,\Phi\, \sqrt{-\frac{1}{2\times 4!}g_{{}_{AE}}\dots
g_{{}_{DH}}\, W^{{}^{ABCD}} \, W^{{}^{EFGH}}\,}\nonumber\\
&&-\frac{1}{4!}\,\int d^{5+1}x \,\sqrt{-g_{(5+1)}}\, W^{{}^{EFGH}}
\partial_{[\, {}_{E}}\, B_{{}_{FGH}\,]}\label{confact}\\
&&\Phi\equiv \epsilon^{A_1\dots A_6}\epsilon_{a_1\dots a_6}
\partial_{A_1}\phi^{a_1}\dots \partial_{A_6}\phi^{a_6}
\end{eqnarray}

where, $\phi^{a_1}$,\dots $\phi^{a_6}$ are six scalar fields
treated as independent degrees of freedom and we consider the
gravitational action in the first order formulation, i.e.
$g_{{}_{AB}}$ and $\Gamma^C_{DE}$ are treated as independent
variables. The connection $\Gamma^C_{DE}$ is torsion-free, i.e.
$\Gamma^C_{DE}=3D\Gamma^C_{ED} $. Thus, $\partial_{[\, {}_{E}}\,
B_{{}_{FGH}\,]}\equiv \nabla_{[\, {}_{E}}\, B_{{}_{FGH}\,]}$ where
$\nabla_{{}_M}$ is the covariant derivative. In Eq.{1}
$R_{{}_{AB}}\equiv R^{{}^C}{}_{{}_{ABC}}$ and
$R^{{}^A}{}_{{}_{BCD}}= \Gamma^{{}^A}_{{}_{B\, C\ , D}}
-\Gamma^{{}^A}{}_{{}_{B\, D\ , C}} + \Gamma^{{}^A}_{{}_{K\,D}}
\Gamma^{{}^K}_{{}_{B\,C}}  - \Gamma^{{}^A}_{{}_{K\,C}} \Gamma^{{}^K}_{{}_{B\,D}}$.\\
$\Phi d^{5+1}x$ is a scalar as well as $\sqrt{-g_{(5+1)}}\,
d^{5+1}x$ under x-coordinates transformation, while under scalar
fields re-definitions:\\

\begin{eqnarray}
&&\phi^{a_j}\longrightarrow \phi^{\prime\,
b_k}\left(\,\phi^{a_j}\,\right)\label{transf}\\
&&\Phi\longrightarrow \Phi^\prime = J\Phi,  J\equiv
\mathrm{det}\left(\, \frac{\partial\phi^{\prime\, a_j}}
{\partial\phi^{b_k} }\,\right)\label{transf2}
\end{eqnarray}

$W^{ABCD}= 3$-brane slope field, it assigns a tangent (hyper)plane
to each spacetime point; the $W$ field is \textit{totally
anti-symmetric} in the four indices.
This field can describe a fluid of {3}-branes [2],[3]\\
$B_{FGH}$ $3$-brane \textit{gauge potential}; the $B$ field is
\textit{totally anti-symmetric} in the three indices. In the last
term the invariant integration measure is written in terms of
$g_{(5+1)}$, instead of $\Phi$ to make the action invariant under
(3). One must in this case assume the
following Weyl rescalings also\\
\begin{eqnarray}
&& g_{{}_{A_1 A_2}}\longrightarrow J\,g_{ {}_{A_1 A_2}} \\
&& g^{{}^{B_1 B_2}}\longrightarrow J^{-1}\,g^{{}^{B_1 B_2}} \\
&& g_{(5+1)}\longrightarrow J^6\,g_{(5+1)} \\
&& W^{{}^{ABCD}}\longrightarrow J^{-3}\,W^{{}^{ABCD}}\\
&& B_{{}_{FGH}}\longrightarrow B_{{}_{FGH}}\ ,\qquad
\Gamma^{{}^A}{}_{{}_{BC}}\longrightarrow \Gamma^{{}^A}{}_{{}_{BC}}
\end{eqnarray}\\
\newpage

Notice that this symmetry holds \textit{only} in the case the
embedding space in $6D$. Let us remark that if we define $W$ as a
``contravariant'' object (upper indices) and $B$ as a covariant
field (lower indices), then the last term in the action $S$
depends on the metric only through $\sqrt{-g_{(5+1)}}$ .\\
We can define the \textit{Dual Representation} of the theory by
changing variables\\

\begin{equation}
W^{{}^{ABCD}} = \frac{1}{2}\frac{\epsilon^{{}^{ABCDEF}}}
{\sqrt{-g_{(5+1)}}}\, \omega_{{}_{EF}}\\
\end{equation}\\

\begin{eqnarray}
S  = -\frac{1}{16\pi G_{(5+1)}}\int d^{5+1}x \, \Phi\, R_{(5+1)}
+ e^2\,\int d^{5+1}x \,\Phi\, \sqrt{\frac{1}{4} g^{{}^{AE}}
g^{{}^{DH}}\, \omega_{{}_{AD}}, \omega_{{}_{EH}}}\nonumber\\
-\frac{1}{6!}\int d^{5+1}x\,\epsilon^{{}^{ABCDEF}} \omega_{{}_{[\,
AB}}\partial_{{}_C}\, B_{{}_{DEF\,]}}\label{sdual}
\end{eqnarray}\\
 
{\bf III Field Equations}\\
We will work out the equations of motion in the dual picture first
and afterwards we will review the brane interpretation of these
solutions.\\
To start let us notice the following facts concerning the
action(11). First it can be written in the form\\

\begin{equation}
S = \int d^{5+1}x \, \Phi\,\left(\, L_G + L_m\,\right)-
\frac{1}{6!} \int d^{5+1}x \,\epsilon^{{}^{ABCDEF}}\, \omega_{
{}_{[\,AB}}\partial_{{}_C}\, B_{{}_{DEF\,]}}
\end{equation}\\

where\\
\begin{eqnarray}
L_G &&= - \frac{1}{ 16\pi G_{(5+1)}}\,g^{{}^{AB}}\,
R_{{}_{AB}}\left(\,\Gamma\,\right)\,\\
L_m &&=e^2\sqrt{\frac{1}{4}\,\omega_{{}_{AB}}\,\omega_{{}_{CD}}\,
g^{{}^{AC}}\, g^{{}^{BD}}}
\end{eqnarray}\\

are homogeneous of degree one in $g^{{}^{AC}}$ , that is\\

\newpage

\begin{equation}
g^{{}^{AB}}\frac{\partial L_m}{\partial g^{{}^{AB}}}=L_m\ ,\qquad
g^{{}^{AB}}\frac{\partial L_G}{\partial g^{{}^{AB}}}=L_G
\end{equation}\\

this property is intimately related to the fact that the action
(11) has the symmetry under $g^{{}^{AB}}\longrightarrow
J^{-1}\, g^{{}^{AB}}$, $\Phi\longrightarrow J\, \Phi $.\\
The equations of motion which result from the variation of the
fields $\phi^a$ are\\

\begin{equation}
\mathbf{A}^{M}_a\,\partial_{{}_M}\,\left(\, L_G + L_m\,\right)=0
\label{phieq}
\end{equation}\\

where\\

\begin{equation}
\mathbf{A}^{M}_m\equiv \epsilon^{{}^{MBCDEF}}\,\epsilon_{mbcdef}\,
\partial_{{}_B} \,\phi^b\,\partial_{{}_C} \,\phi^c\,\partial_{{}_D}\,
\phi^d\,\partial_{{}_E} \,\phi^e\,\partial_{{}_F} \,\phi^f
\end{equation}\\

Since $\mathrm{det}\left(\,
\mathbf{A}^{M}_m\,\right)=6^{-6}\Phi^6/6!$ Then we have that if
$\Phi\ne 0$, this means that (16)implies
\begin{equation}
L_G + L_m=M=\mathrm{const}.
\end{equation}
The equation of motion obtained from the variation of
$g^{{}^{AB}}$ is
\begin{equation}
-\frac{1}{16\pi\, G_{(5+1)}}\, R_{{}_{AB}}+
\frac{\partial\,L_m}{\partial g^{{}^{AB}}}=0 \label{einsteq}
\end{equation}
by contracting (19) with respect to $g^{{}^{AB}}$ and using the
homogeneity property of $L_m$, we obtain that the constant of
integration $M$ equals zero. Evaluating (19) we find
\begin{equation}
R_{{}_{AB}}=4\pi\, e^2\, G_{(5+1)}\frac{\omega_{{}_{AC}}\,
\omega_{{}_B}{}^{{}^C}}
{\sqrt{\frac{1}{4}\,\omega_{{}_{MN}}\,\omega^{{}^{MN}}\, }}
\label{}
\end{equation}

Eq.(20) is also consistent with the Einstein form

\begin{eqnarray}
&& R_{{}_{AB}}-\frac{1}{2}g_{{}_{AB}}\,R= -8\pi G_{(5+1)} \,
T_{{}_{AB}} \label{rt}\\
&& T_{{}_{AB}}= -2\frac{\partial L_m}{\partial g^{{}^{AB}}}
+g_{{}_{AB}}\, L_m
\end{eqnarray}
which for $L_m$ is given by
\begin{equation}
T_{{}_{AB}}=\frac{e^2}{2}\frac{ \omega_{{}_{AC}}\,  \omega_{
{}_B}{}^{{}^C}}
{\sqrt{\frac{1}{4}\,\omega_{{}_{MN}}\,\omega^{{}^{MN}}} }-e^2\,
g_{{}_{AB}}\sqrt{\frac{1}{4}\,\omega_{{}_{MN}}\,
\omega^{{}^{MN}}}\label{tab}
\end{equation}
\newpage
as one can easily check that solving from $R$ by contracting both
sides of (21) with $T_{{}_{AB}}$ given by (23) and then
replacing $R$ into (21) gives (19).\\
Let us consider now the equation of motion for the connection
coefficients $\Gamma^{{}^A}{}_{{}_{BC}}$. Defining
\begin{equation}
\bar g_{{}_{AB}}=\left(\, \frac{\Phi}{\sqrt{-g_{(5+1)}}}\,
\right)^{1/2}\, g_{{}_{AB}} \label{gconf}
\end{equation}

one can verify that
\begin{equation}
\Phi\, g^{{}^{AB}}=\sqrt{-\bar g_{(5+1)}}\, \bar g^{{}^{AB}}
\end{equation}

Therefore, the equation of motion for $\Gamma^{{}^A}{}_{{}_{BC}}$
is obtained by the condition that the functional

\begin{equation}
I\equiv -\frac{1}{16\pi\,\pi G_{(5+1)}} \int d^{5+1}x \,
\sqrt{-\bar g_{(5+1)}} \,\bar g^{{}^{AB}}\,
R_{{}_{AB}}\left(\,\Gamma\,\right)
\end{equation}
is extremized under variation of $\Gamma^{{}^A}{}_{{}_{BC}}$. This
is however the well known Palatini problem in General Relativity
(~but where the metric $\bar g_{{}_{AB}}$ enters, not the original
metric $g_{{}_{AB}}$ ~). Therefore $\Gamma^{{}^A}{}_{{}_{BC}}$ is
the well known Christoffel symbol, but not of the metric
$g_{{}_{AB}}$ rather than the metric $\bar g_{{}_{AB}}$:

\begin{equation}
\Gamma^{{}^A}{}_{{}_{BC}}=\left\{\,{}_B {}^A {}_C\,\right\}
\vert_{\bar g}
\end{equation}\\

Notice the interesting fact that $\bar g_{{}_{AB}}$ is conformally
invariant, i.e. invariant under the set of transformations (3),
(4),(5).\\
Also, in the gauge $\Phi=\sqrt{-g_{(5+1)}}$, the metric
$g_{{}_{AB}}$equals the metric $\bar g_{{}_{AB}}$ so one may call
this the ``Einstein gauge'', since here all non-Riemannian
contributions to the connection disappear. Alternatively, without
need of choosing a gauge one may choose to work with the
conformally invariant metric $\bar g_{{}_{AB}}$ in terms of which
the connection equals the Christoffel symbol and all
non-Riemannian structures disappear. Finally the equations of
motion obtained from the variation of the gauge fields
$\omega_{{}_{AB}}$ and $B_{{}_{MNP}}$ are
\begin{equation}
\Phi\,\frac{\omega^{{}^{AB}}}{\sqrt{\frac{1}{4}\omega_{{}_{MN}}
\omega^{{}^{MN}}}}=
\frac{1}{6!}\epsilon^{{}^{ABCDEF}}\partial_{[\, {}_{C}}\,
B_{{}_{DEF}\,]} \label{fieq}
\end{equation}

and\\

\begin{equation}
\epsilon^{{}^{ABCDEF}}\partial_{[\, {}_{D}}\,
\omega_{{}_{EF}\,]}=0 \label{curl}
\end{equation}
taking the divergence of (28) we obtain\\

\begin{equation}
\partial_{{}_A} \left(\,
\Phi\,\frac{\omega^{{}^{AB}}}{\sqrt{\frac{1}{4}\omega_{{}_{MN}}
\omega^{{}^{MN}}}}\, \right)=0 \label{diveq}
\end{equation}
\newpage

{\bf IV. Brane-world solutions in the Dual Picture}\\

In this section we are going to consider the product spacetime

\begin{equation}
ds^2=g_{\mu\nu}(x_{||})\, dx^\mu_{||}\, dx^\nu_{||}+
\gamma_{ij}\left(\, \vec x_\perp\,\right)\, dx_\perp^i\,
dx_\perp^j \label{lelem}
\end{equation}\\

where $\mu\ ,\nu=0\ , 1\ , 2\ , 3$ and $i\ ,j= 4\ , 5$.
Furthermore, we consider a slope field $W^{{}^{ABCD}}$ with
non-vanishing components only in the first four coordinates
$\left(\, 0\ ,1\ ,2\ ,3\,\right)$, which means we are dealing with
a set of parallel branes orthogonal to the extra-dimensions, if we
use the brane interpretation requested of the field refs. [2],[3].
This means that the dual field $\omega_{AB}$ has non-zero
components in the 4 ,5 directions only. In this case, we see from
eq.(19) that the Ricci curvature induced in the four dimension $0\
,1\ ,2\ ,3$, is zero:

\begin{equation}
R_{\mu\nu}=0\label{ricci0}
\end{equation}\\

Thus, the ordinary four dimensions (accessible to our experience)
are not curved by this kind of matter. This is a very important
remark, since there is no need to introduce a bare cosmological
constant to cancel some contribution from the gauge field,
\textit{no type of fine  tuning},  most usual in extra dimensional
theories, \textit{is needed here}.\\
The simplest solution of (32) is \textit{flat}, four dimensional
spacetime

\begin{equation}
g_{\mu\nu}=\eta_{\mu\nu}\\
\end{equation}

Let us analyze now the additional field equations. It is
convenient to choose gauge $\Phi=\sqrt{-g_{(5+1)}}$, even if the
conformally invariant metric $\bar g_{{}_{AB}}$ gives the same results.\\
The two-dimensional metric $\gamma_{ij}$ can always be put in a
conformally flat form, i.e. one can always choose a  coordinate
system where

\begin{equation}
\gamma_{ij}\, dx_\perp^i\, dx_\perp^j= \psi\left(\, x^4\ , x^5\,
\right)\left[\, \left(\, dx^4\,\right)^2 + \left(\,
dx^5\,\right)^2\,\right] \label{2metric}\\
\end{equation}

As far as the dual slope field is concerned, its most general form
along the extra dimension where it is non-zero, is dictated by its
tensorial structure in two-dimensions, which is\\

\begin{equation}
\omega^{ij}= -\frac{\epsilon^{ij}}{\sqrt\gamma} \rho\left(\, x^4\
, x^5\,\right)\ , \qquad \gamma\equiv
\mathrm{det}\left(\,\gamma_{ij}\,\right) \label{omega}
\end{equation}\\

It turns out that the field equations do not determine the
function $\rho$ as

\begin{equation}
\partial_i\left(\, \frac{\omega^{ij}\,\sqrt{\gamma}}{\sqrt{-\frac{1}{2}
\omega^{kl}\, \omega_{kl}}}\,\right)=0\longrightarrow
\partial_i\epsilon^{ij}=0
\end{equation}
\newpage

which is ``trivially'' satisfied $\epsilon^{ij}$ being the totally
anti-symmetric symbol in two-dimensions.\\
The function $\rho\left(\,  x^4\ , x^5\,\right)$ acts , however,
as a source that determines the metric. The physical source of the
arbitrariness in $\rho$ can be understood by invoking the brane
interpretation of the $\omega$-field. The function $\rho$ is
associated to the density of $3$-branes being piled in the extra
dimensions ref [4].  Since these branes do not exert any force one
upon each other they can be accumulated with an arbitrary density
at each extra-dimensional point. Recalling that the scalar
curvature of (\ref{2metric}) is $R=-\psi^{-1}\nabla^2\psi$, we
have from $R=16\pi G_{(5+1)}L_m$:\\

\begin{equation}
-\frac{1}{\psi}\nabla^2 \psi=16\pi G_{(5+1)} \,\rho
\label{metrica}
\end{equation}

$\rho$ is free to be taken any possible values, but once it is
assigned $\psi$ is determined by (\ref{metrica}). The argument can
be also reversed: for any $\psi$ (\ref{metrica}) gives the
corresponding $\rho$. An interesting case is obtained when rho
consists of a constant part plus one or more delta function parts.
Since $R$ is a scalar a delta function part can appear only in
combination $\delta^{(2)}/\sqrt{\gamma}$. Let us define:

\begin{eqnarray}
r&&=\sqrt{  (x^4)^2 + (x^5)^2 }\label{s2}\\
x^4&&= r\sin\phi\label{s3}\\
x^5&&= r\cos\phi\label{s4}
\end{eqnarray}\\

which describe the metric close to $r=0$, and take
$\psi=\psi(r)$, so\\

\begin{equation}
\gamma_{ij} dx_\perp^i\, dx_\perp^j= \psi(r)\left(\, dr^2 + r^2
d\phi^2\,\right)
\end{equation}\\

Then, using the representation of the delta-function (with
integration measure $r d\phi dr $)\\

\begin{equation}
\delta^{(2)}\left(\, r\, \right)= \frac{1}{2\pi}\nabla^2\ln r
\end{equation}\\

where $\nabla^2=\frac{d^2}{dr^2}+ \frac{1}{r}\,\frac{d}{dr}$.
Then, for\\

\begin{equation}
\rho=\sqrt 2\, B_0 + T\frac{ \delta^{(2)}\left(\, r\,
\right)}{\psi } \label{rho}
\end{equation}
\newpage
where $B_0$ and $T$ are constants. By inserting (43) into (37) we
obtain (similar equation was obtained in ref [20] in the context
of $2+1$ gravity)\\

\begin{equation}
\psi=\frac{4\alpha^2 b^2}{r^2}\left[\, \left(\, \frac{r}{r_0}
\right)^\alpha +  \left(\, \frac{r}{r_0}
\right)^{-\alpha}\,\right]^{-2}
\end{equation}\\

where

\begin{eqnarray}
&&\alpha \equiv 1-4 G_{(5+1)}T\\
&& b^2\equiv \frac{\sqrt 2}{16\pi G_{(5+1)}B_0}
\end{eqnarray}

Such a metric can be transformed into the form\\

\begin{equation}
\gamma_{ij}\, dx_\perp^i\, dx_\perp^j=b^2\,\left(\, d\theta^2 +
\alpha^2\,\sin^2\theta\, d\phi^2\,\right)
\end{equation}

where $\phi$ ranges from $0$ to $2\pi$, or, equivalently,

\begin{equation}
\gamma_{ij}\, dx_\perp^i \, dx_\perp^j= b^2\,\left(\, d\theta^2 +
\sin^2\theta\, d\bar\phi^2\,\right)
\end{equation}\\

where $\bar\phi$ ranges from $0$ to $2\alpha\pi< 2\pi$. A complete
solution must contain two branes (in the coordinate system
(38),(39),(40) we are able  to display only one pole of the sphere
, the other one is at the other pole of the sphere, where in $(r\
,\phi)$ coordinates is at $r\to\infty$). Here the term
``branes'' means delta-functions contributions to $\rho$.\\
Of course, this solution is one out of a continuum of solutions,
but is interesting because it allows us to connect to other works
on the subject ref. [17] where similar effects are discussed.\\
Nevertheless, we stress the fact that the function $\rho$ is
totally free in this conformally invariant model. The situation
can change once conformal breaking contributions are allowed. This
will be the subject of the next section.\\
\newpage
{\bf V. The Introduction of Conformal Symmetry Breaking}\\

We consider now a generalization of (1),(2),\\
\begin{eqnarray}
S=\int d^{5+1}x \, \Phi\, \left(\, -\frac{1}{16\pi
G_{(5+1)}}g^{{}^{AB}}\, R_{{}_{AB}}\left(\, \Gamma\,\right) +
\frac{1}{2}g^{{}^{AB}} \partial_{A}\alpha \partial_{B}\alpha
- V(\alpha) \,\right)  \nonumber\\
+\,\int d^{5+1}x \,\Phi\, F \left(\, \sqrt{-\frac{1}{2\times
4!}g_{{}_{AE}}\dots g_{{}_{DH}}\, W^{{}^{ABCD}} \,
W^{{}^{EFGH}}\,}  \,\right)\\
-\frac{1}{4!}\,\int d^{5+1}x \,\sqrt{-g_{(5+1)}}\, W^{{}^{EFGH}}
\partial_{[\, {}_{E}}\, B_{{}_{FGH}\,]}\label{confact-1}\\
\Phi\equiv \epsilon^{A_1\dots A_6}\epsilon_{a_1\dots a_6}
\partial_{A_1}\phi^{a_1}\dots \partial_{A_6}\phi^{a_6}\nonumber
\end{eqnarray}

We have now introduced a new (in principle) degree of freedom, the
scalar field $\alpha$. In the above expression, conformal symmetry
is broken in two different ways: i) by the potential $V$ of the
scalar field $\omega$ and ii) by the introduction of a function F,
which gives rise to conformal symmetry breaking in the case this
function is a non linear function of its argument.\\

Once again, going to the dual picture\\

\begin{equation}
W^{{}^{ABCD}} = \frac{1}{2}\frac{\epsilon^{{}^{ABCDEF}}}
{\sqrt{-g_{(5+1)}}}\, \omega_{{}_{EF}} \label{dual1}
\end{equation}
We obtain now,\\
\begin{eqnarray}
S=\int d^{5+1}x \, \Phi\,\left(\, -\frac{1}{16\pi
G_{(5+1)}}R_{(5+1)} + \frac{1}{2}g^{{}^{AB}}
\partial_{A}\alpha\partial_{B}\alpha - V(\alpha) \,\right) +
\nonumber\\
\int d^{5+1}x \,\Phi\, F \left(\,\sqrt{\frac{1}{4}\,
g^{{}^{AE}} g^{{}^{DH}}\, \omega_{{}_{AD}},
\omega_{{}_{EH}}}\,\right) -\frac{1}{6!}\int
d^{5+1}x\,\epsilon^{{}^{ABCDEF}} \omega_{{}_{[\,
A}}\partial_{{}_C}\, B_{{}_{DEF\,]}}\label{sdunl}
\end{eqnarray}\\

{\bf VI. Curvature and a Constraint Equation in the case
of Conformal\\ Symmetry Breaking}\\

The equations of motion which result from the variation of the
fields $\phi^a$ are

\begin{equation}
\mathbf{A}^{M}_a\,\partial_{{}_M}\,\left(\, L_G + L_m\,\right)=0
\label{phenl}
\end{equation}\\

where

\begin{equation}
\mathbf{A}^{M}_m\equiv \epsilon^{{}^{MBCDEF}}
\,\epsilon_{mbcdef}\,
\partial_{{}_B} \,\phi^b\,\partial_{{}_C} \,\phi^c\,\partial_{{}_D}\,
\phi^d\,\partial_{{}_E} \,\phi^e\,\partial_{{}_F} \,\phi^f\\
\end{equation}\\

\newpage
Since $\mathrm{det}\left(\,
\mathbf{A}^{M}_m\,\right)=6^{-6}\Phi^6/6!$. Then we have that if
$\Phi\ne 0$, this means that (53) implies

\begin{equation}
L_G + L_m=M=\mathrm{const}.
\end{equation}

As opposed to the conformally invariant case, the constant $M$
will not be determined to be zero, but can in principle remain
undetermined. In fact, it could play a very important role in the
spontaneous symmetry breaking of internal symmetries.\\

The equation of motion obtained from the variation of $g^{AB}$ gives
us the curvature equation:

\begin{equation}
-\frac{1}{16\pi\, G_{(5+1)}}\, R_{{}_{AB}}+
\frac{\partial\,L_m}{\partial g^{{}^{AB}}}=0 \label{eineqnl}
\end{equation}\\

by contracting (56) with respect to $g^{{}^{AB}}$ , using also eq.
(55), we get that\\

\begin{equation}
g^{{}^{AB}}\frac{\partial L_m}{\partial g^{{}^{AB}}}=L_m\ - M
\end{equation}

It is very important to notice that since $L_m$ is not an
homogeneous function of degree one of $g^{{}^{AB}}$, then, as
anticipated, $M$ will not necessarily vanish. As it is apparent,
all non homogeneous of degree one pieces of the Lagrangian will
enter into the above equation, these are exactli the conformal
breaking terms. When inserting our specific lagrangian density, we
obtain a very interesting constraint equation:
\begin{equation}
u \frac {dF(u)}{du} - F(u) +V+M = 0
 \end{equation}

where

\begin{equation}
u = \sqrt{\frac{1}{4} g^{AE} g^{DH}
\omega_{{}_{AD}} \omega_{{}_{EH}}}
\end{equation}\\

{\bf VII Zero 4-D Cosmological Constant and The fixed or "D-Brane"\\
boundary conditions for the scalar field on the singular branes}\\

After the study of the curvature equation and the constraint
equation in the previous section, we are now in conditions to
discuss the basic effects associated with the model with breaking
of conformal invariance. In order to do so, it is useful to
understand the meaning of the function $F$. In the situation of
conformal invariance, it is fundamental that the function F be a
linear function. Even in the situation of scale invariance
breaking, if we still wish to have as our solution, the singular
brane case, then we must have that as u becomes very large, then
$F(u)$ $ \rightarrow$ $ C $ $u $, where $C$ is a constant. The
linearity of\\
\newpage
$F$ in $u$, or what is the same, the square root choice (in eq. 1 or
in eqs. 12-14) is a requisite for the existence of singular brane
(see also refs [2] and [3]). Then, the construction of the singular
branes studied in section IV of the paper becomes possible. For
example we get in this limit eq. (36) that a singularity in
$\omega$  is cancelled. In this limit we clearly see that
$u \frac {dF(u)}{du}$ - F(u)= 0, which by equation (58) means
that at the location of the singular brane, the scalar field
is "frozen " at the values determined by\\

\begin{equation}
V+M = 0
\end{equation}
Therefore the brane acts like a "D-Brane", giving the scalar field
a boundary condition and in general an expectation value, or a non
trivial average value in the bulk as well. This even without
introducing tachyonic mass terms in general.\\
Now, concerning the question of the vanishing of the 4-D
cosmological constant: If we just make the assumption that the
compactified solutions are such that the field strength
$\omega_{AB}$ gets an expectation value for the values of $A$ and
$B$ in the extra dimensions only, and that the scalar field
$\alpha$ has non trivial gradient only in the extra dimensions,
then as a consequence, the matter lagrangian does not depend on
the metric components  $g_{\mu \nu }$ and eq. (56) tells us then
immediately that\\

\begin{equation}
R_{\mu \nu } = 0,  (\mu, \nu  = 0, 1, 2, 3)
\end{equation}

That is the four dimensional part of the manifold once again does
not get curved.\\

{\bf X The complete system of equations}\\

We have already discussed some equations of motion, in fact the
equations of motion which are responsible for the most important
effects, that is that only the extra dimensions aquire curvature
and that the scalar field $\alpha$ is fixed at some value
determined by an arbitrary constant of integration $M$. Those
conclusions do not depend on the details of the solutions, but on
very general features. A complete discussion requires however the
consideration of all the equations of motion. \\

The equation for the connections gives the solution that the
connections are the Christoffel symbols of a conformally
transformed metric, exactly as it was in the conformally invariant
case, i.e. the connections are the Christoffel symbols of the
metric\\

\begin{equation}
\bar g_{{}_{AB}}=\left(\, \frac{\Phi}{\sqrt{-g_{(5+1)}}}\,
\right)^{1/2}\, g_{{}_{AB}} \label{gconf-1}
\end{equation}
Now,  as opposed to the conformally invariant case, one does not
have the freedom to choose that factor to be one. Since the
Riemannian $4D$ curvature of the barred metric is zero, we have
the right to consider solutions where
\newpage
\begin{equation}
\bar g_{\mu\nu}=\eta_{\mu\nu}\ .
\end{equation}

while the extra dimensional part of the metric can (and in general
must)be curved. Let us define the following notation,

\begin{equation}
\bar g_{ij}=\bar \gamma_{ij}\ ,   g_{ij}= \gamma_{ij}
\end{equation}
where of course the relation of the extra dimensional bar and
unbarred metrics is\\
\begin{equation}
\bar \gamma_{ij} \ = \left(\, \frac{\Phi}{\sqrt{-g_{(5+1)}}}\,
\right)^{1/2}\, \gamma_{ij}
\end{equation}

here $i,j = 4,5$.  The gauge field equation (expressed in terms
of the original metric, not the barred one) is\\

\begin{equation}
\partial_{A}\ \left(\
\frac{dF}{du}\Phi\frac{\omega^{AB}}{\sqrt{\omega_{CD}\omega^{CD}}}\right)=0
\end{equation}\\

Assuming the $\omega_{AB}$ to have expectation value only for
$A,B$ $=$  $i,j = 4,5$, we obtain that $\omega_{ij}$ has only one
independent component, because of the antisymmetry of such tensor,
then in equation (66) the determinant of the internal
metric appears. Such equation can be integrated to give \\

\begin{equation}
\frac{dF}{du} \frac {\Phi}{\sqrt{\gamma}} = C
\end{equation}

where $C$ is some constant. The above equation allows us to
determine the measure $\Phi$ in terms of $\sqrt{\gamma}$ and $u$,
while $u$ itself is determined in terms of the scalar field
through eq. (58), except at the singular branes, when u is absent
from such equation. This equation becomes instead an equation
which fixes the value of the scalar field $\alpha$ at such
boundaries, that is, it gives a Dirichlet boundary condition for the scalar field.\\
The $i,j$ components of the gravitational equations are\\

\begin{equation}
\frac{1}{16\pi G_{(5+1)}}R_{ij} = \frac{1}{2} u\frac{dF}{du}
\gamma_{ij}+ \frac{1}{2}\partial_{i}\alpha \partial_{j}\alpha
\end{equation}\\

All the system of equations appears then well defined. The two
basic features that we have focused on, that is the fact that the
ordinary dimensions do not get curved and the fact that the scalar
field gets fixed at the bounday, do not depend on the details.
Only on the fact that the scalar field has gradients only in the
extra dimensions, which is certainly consistent since the boundary
conditions are fixed by the branes, which have positions in the
extra dimensions but which are totally homegeneous with respect to
the $4$ dimensions. The other assumption is of course that the
gauge field $\omega_{AB}$ gets vacuum expectation value in the
extra dimensions only.
\newpage
All details of the specifics of the solution are not going to
change these facts, although they may be very important for the
phenomenology of the theory. For example one may study how the
scalar field $\alpha$ interpolates between the value determined by
the boundary condition at the brane, i.e $V+M=0$ and a value close
to the minimum of the potential V may be close to the middle
region between the two branes. \\
The introduction of conformal breaking terms can be discussed in
the context of "degenerate perturbation theory": When conformal
symmetry is present the four dimensional space is flat, while the
extra dimensional part is largelly arbitrary. This is because in
the conformal symmetric case in the continious distribution of
branes, the branes do not interact with each other. Any particular
brane does not suffer any force from the others  and therefore
those branes can be pilled with an arbitrary density in the extra
dimension.\\ This large degeneracy is broken once conformal
symmetry is broken, one specific profile or density of branes
appears singled out as the solution. \\

{\bf X Discussion and conclusions}\\

In this paper we have discussed how the gauge formulation of
branes can be used in the framework of ``brane world'' scenarios.\\
The formulation of $3$-branes in a six-dimensional target
spacetime can be made in a conformally invariant way. This is
possible for extended objects in case the target spacetime has two
more dimensions than the extended object itself. \\
This conformal invariance is intimately related to fact that the
branes (or equivalently the associated gauge fields) only curve
the manifold orthogonal to the brane, the extra-dimensions. No
fine tuning of a $6D$ cosmological constant is needed in this
case. Therefore, no ``old cosmological constant problem'' , as
Weinberg has defined it [21], appears.\\
An interesting phenomenon is that the parallel $3$-branes can be
found with an arbitrary density for any value of $\vec
x_\perp=\left(\, x^4\ , x^5\,\right)$. The density $\rho\left(\,
\vec x_\perp\,\right)$ cannot be determined. This represents a
large degeneracy and, therefore, a freedom in the possible
ways the branes can be accounted in the extra dimensions.\\
The basic feature, that the matter curves only the extra
dimensions is related to the fact that one is able to formulate
the theory in terms of the measure $\Phi$, since then Eq.(19)
follows automatically. Provided we adopt such formulation Eq.(19)
tell us that if $L_m$ depends only from $\gamma_{ij}$, then only
extra dimensions are curved. Conformal invariance holds if 
the embedding space is $6D$.\\

We then generalize to include conformal breaking terms. The terms
which we include and which break the conformal invariance are of
two types: one, which introduces a "non ideal" fluid behavior for
the branes, but leaves singular branes solutions unchanged and the
other conformal breaking term is a scalar field potential. At the
singular brane, the scalar field gets frozen at an expectation
value determined by the eq. $V+M=0$. This can be a mechanism that
could introduce spontaneous breaking of internal symmetries,
due to boundary conditions. Breaking of internal symmetries\\
\newpage
by boundary conditions has been studied recently by several
authors [22], although these authors use this effects to advocate
the possibility of spontaneous symmetry breaking without a Higgs
field. In our case, the existence of a scalar (that is a Higgs)
still appears necessary, but the spontaneous symmetry breaking can
be achieved by means of the boundary conditions at the wall
$V+M=0$, which will require a condensation of the scalar field at
the branes, irrespective of the existence of a tachyonic mass term
in $V$. Under very general conditions, the solutions do not curve
the four dimensional space, but only the extra dimensional
space. \\

{\bf Acknowledgments}. \\
I want to thank A.Kaganovich, E.Nissimov, S. Pacheva and E. Spallucci for very helpful
conversations. I would also like to thank the Bulgarian Academy of Sciences, 
the Department of
Theoretical Physics of the University of Trieste, I.N.F.N. and the
University of Michigan for the hospitality and support.


\end{document}